\documentclass[10pt,twocolumn,letterpaper]{article}

\usepackage{styles/cvpr}
\usepackage{times}
\usepackage{epsfig}
\usepackage{graphicx}
\usepackage{amsmath}
\usepackage{amssymb}
\usepackage{tabularx}
\usepackage{url}
\usepackage{todonotes}
\usepackage{multirow}
\usepackage{subfigure}

\usepackage[pagebackref=true,breaklinks=true,letterpaper=true,colorlinks,bookmarks=false]{hyperref}

\cvprfinalcopy 

\def\cvprPaperID{****} 

\ifcvprfinal\fi
\begin{document}

\title{Biometric Template Storage with Blockchain: \\A First Look into Cost
and Performance Tradeoffs}

\author{Oscar Delgado-Mohatar, Julian Fierrez, Ruben Tolosana  and Ruben Vera-Rodriguez\\
Escuela Politecnica Superior\\ Universidad Autonoma de Madrid, Madrid, Spain\\
{\tt\small \{oscar.delgado, julian.fierrez, ruben.tolosana, ruben.vera\}@uam.es}}



\maketitle

\begin{abstract}
   We explore practical tradeoffs in blockchain-based biometric template
storage. We first discuss opportunities and challenges in the integration of
blockchain and biometrics, with emphasis in biometric template storage and protection, a
key problem in biometrics still largely unsolved. Blockchain technologies provide
excellent architectures and practical tools for securing and managing the
sensitive and private data stored in biometric templates, but at a cost. We
explore experimentally the key tradeoffs involved in that integration, namely:
latency, processing time, economic cost, and biometric performance. We
experimentally study those factors by implementing a smart contract on Ethereum for biometric template storage,\footnote{Deployed to the Ethereum Ropsten testnet at address: {\small\texttt{0x8f737f448de451db9b1c046be7df3b48839673a1}}} whose cost-performance is evaluated by varying the complexity of state-of-the-art
schemes for face and handwritten signature biometrics. We report our
experiments using popular benchmarks in biometrics research, including deep learning approaches and databases captured in the wild. As a result, we experimentally show that straightforward schemes for data storage in blockchain (i.e., direct and hash-based) may be prohibitive for biometric template storage using state-of-the-art biometric methods. A good cost-performance tradeoff is shown by using a blockchain approach based on Merkle trees.
\end{abstract}

\section{Introduction}

The integration of the advantages and characteristics of public blockchains in biometric systems is a very recent area of research, but with a high potential and interest.

Combining blockchain and biometrics could potentially have many advantages. As a first approximation, the blockchain technology \cite{nanda19block} could provide biometric systems with some desirable characteristics such as \textbf{immutability}, \textbf{accountability}, \textbf{availability} or \textbf{universal access}. These properties enabled by blockchain technology may be very useful, among other applications in biometrics, to secure the biometric templates \cite{nanda15template}, and to assure privacy in biometric systems \cite{bringer13privacy}.

However, despite these opportunities, the current blockchain technology suffers from some potential limitations that must be carefully studied and characterized before the combination of both biometrics and blockchain technologies. 

The main contribution of this study is two fold: 1) we analyze cost and performance tradeoffs
when using blockchain for biometric template storage. We first discuss the existing alternatives for the storage of large volumes of data in blockchains, and how the complexity of schemes for face and handwritten signature biometrics affects to the cost and execution time of the final system; and 2) we experimentally measure these factors, optimizing the storage requirements of each biometric scheme while keeping their performances.


The remainder of the paper is organized as follows. In Section \ref{sec:blockchain_for_biometrics} a description of the most relevant features of blockchains, and the challenges and limitations of the technology that directly affect to biometric technologies is provided. In Section \ref{sec:storage_analysis}, we describe three popular storage techniques for public blockchains, briefly analyzing their main characteristics. Sections \ref{sec:methods} and \ref{sec:protocol_setup} present the setup and methods used in the experiments, whose results are shown in Section \ref{sec:results}. Finally, Section \ref{sec:conclusions} draws the final conclusions.

\section{Blockchain for Biometrics}\label{sec:blockchain_for_biometrics}

\subsection{Smart contracts}\label{subsec:applications}

A \textit{smart contract} is, essentially, a piece of code executed in a secure environment that controls digital assets. Examples of these secure environments include regular servers controlled by ``trusted parties'', decentralized networks (blockchains), or servers with secure hardware (SGX) \cite{Karande2017,Kucuk2016}.

Many public blockchains support the execution of smart contracts, but Ethereum \cite{Dannen2017} is currently considered the most reliable, secure and used. In essence, Ethereum could be seen as a distributed computer, with capability to execute programs written in Turing-complete, high-level programming languages. These programs comprise a collection of pre-defined instructions and data that has been recorded at a specific address of a blockchain. For biometric purposes, a smart contract running in a blockchain can assure a semantically correct execution. 


\subsection{Challenges and limitations}\label{subsec:challenges}
Despite the new opportunities already described in previous sections, the combination of both blockchain and biometric technologies is not straightforward due to the limitations of the current blockchain technology. Among them, it is important to remark: 1) its transaction processing capacity is currently very low (around tens of transactions per second), 2) its actual design implies that all system transactions must be stored, which makes the storage space necessary for its management to grow very quickly, and 3) its robustness against different types of attacks has not been sufficiently studied yet. 

In addition, public blockchains suffer from other limitations which could impact the deployment and integration with biometric systems:

\begin{itemize}
\label{sec:coste}
\item \textbf{Economic cost of executing smart contracts:} In order to support smart contracts in blockchains (like Ethereum), and to reward the nodes that use their computing capacity to maintain the whole system, each instruction executed requires the payment of a fee in \textit{gas} units. This gas is paid in the native cryptocurrency of Ethereum, called \textit{ether}.

\item \textbf{Privacy:} By design, all operations carried out in a public blockchain are known to all the participating nodes. Thus, it is not possible to directly use secret cryptographic keys, which reduces the number of potential applications. 


\item \textbf{Processing capability:} Another important limitation is related to its processing capability. Ethereum, for example, is able to run just around a dozen transactions per second, what it could be not enough for some scenarios. 


\item \textbf{Scalability:} Currently, the size of the public blockchains (Bitcoin and Ethereum) is around 200GB, and it is growing very fast. This can be a problem for some application scenarios such as the Internet of Things (IoT).

\end{itemize}



\section{Storage requirements analysis}\label{sec:storage_analysis}

As stated in the previous section, one of the main potential limitations for the integration of both technologies is the cost of running a biometric system (totally or partially) in a blockchain. It is therefore crucial to properly estimate and minimize that cost. The present paper is an initial attempt in that regard.

This section describes the different existing schemes to store large volumes of data (e.g., a database of biometric templates) in public blockchains with smart contracts execution capabilities, like Ethereum.

There are essentially three approaches, which are presented below in terms of complexity (from lower to higher), and economic cost (from higher lo lower):

\begin{itemize}
    \item \textbf{Full on-chain storage}: all data is stored, as-is, in the blockchain. 
    \item \textbf{Data hashing}: the blockchain only stores a hash of the data that guarantees its immutability. The data itself is stored off-chain in other system: distributed (e.g., IPFS \cite{Benet2014}), cloud, or local.
    \item \textbf{Merkle trees}: data is stored also off-chain, but it is preprocessed by constructing a Merkle tree structure, which reduces storage costs and increases the bandwidth.
\end{itemize}

These alternatives are discussed in more detail next.

\subsection{Full on-chain storage}

This is the simplest scheme and therefore, the most inefficient and costly. In this case, the data are just stored in the blockchain as is, without any type of pre-processing. For example, biometric templates could be directly stored as a data structure in a smart contract, as part of a more general digital identity model. 


In general terms, the storage space in public blockchains is specially expensive compared to computation, in order to discourage its abusive use. Therefore, as shown by experiments and figures presented in Section \ref{sec:results}, the use of this storage scheme would commonly imply a prohibitive cost for most biometric applications.


As an example, Table \ref{tab:full-on-chain} depicts the cost of reading and storing 1 Kilobyte of data in Ethereum in terms of gas units, ether, and US dollars.

    \begin{table}[]
    \centering
        \begin{tabular}{cccc}
        \hline
        \multicolumn{1}{|c|}{\textbf{Operation}} & \multicolumn{1}{c|}{\textbf{Gas/KB}} & \multicolumn{1}{c|}{\textbf{ETH/KB}} & \multicolumn{1}{c|}{\textbf{\$/KB}} \\ \hline
        \multicolumn{1}{|c|}{READ}               & \multicolumn{1}{c|}{6,400}           & \multicolumn{1}{c|}{0.000032}        & \multicolumn{1}{c|}{\$0.004}         \\ \hline
        \multicolumn{1}{|c|}{WRITE}              & \multicolumn{1}{c|}{640,000}         & \multicolumn{1}{c|}{0.0032}          & \multicolumn{1}{c|}{\$0.448}         \\ \hline
        \multicolumn{1}{l}{}                     & \multicolumn{1}{l}{}                 & \multicolumn{1}{l}{}                 & \multicolumn{1}{l}{}               
        \end{tabular}
    \caption{Non-volatile storage costs in Ethereum. We have considered a gas price of 1 gwei (1 gwei = $10^{-9}$ ETH), and 1 ETH = \$140 (at time of writing, March 2019).}\label{tab:full-on-chain}
    \end{table}


\subsection{Data hashing}

To overcome the problems of the previous scheme, a more efficient approach is to store the data \textit{off-chain} and use the blockchain just as a integrity guarantee due to its intrinsic immutability. This way, instead of the full data, only a hash value of it is stored in the blockchain (smart contract). Then, the complete template can be stored in any other traditional external storage system (see Figure \ref{fig:hash_merkle_tree}).

This possibility provides a great flexibility, because any platform, as public clouds or existing corporate servers, can be used to store the full set of biometric templates. In any case, to maintain the distributed spirit, resistance to censorship and high availability of public blockchains, distributed storage systems such as IPFS or Swarm \cite{Ozyilmaz2018} would be desirable in this case. 

On the other hand, this approach can make use of any cryptographic hash function, such as the SHA3 family, which can produce outputs from 224 to 512 bits in length \cite{sha3}. In this work, we consider hashes of 256 bits per template, which, in any case, can greatly reduce storage costs compared to full on-chain storage.

One drawback of this approach is that it is still necessary to ensure the availability of the data stored outside the blockchain. If these data were lost or tampered, even when this modification would be always noticed, the viability of the system would be compromised.


\subsection{Merkle trees}

Finally, the previous scheme can be still further improved, through the use of a data structure known as \textit{Merkle tree} \cite{Merkle1987}. This construction is widely used in cryptography and computer science problems such as database integrity verification \cite{Mouratidis2009}, peer-to-peer networks \cite{5452454} and, of course, blockchains \cite{Dannen2017}.

A Merkle tree is a binary tree data structure in which every node contains the cryptographic hash of the concatenation of its child nodes contents. Due to this recursive way of constructing itself, the tree root contains statistical information of the rest of nodes, and the modification of any node content will cause the complete change of the value of the root. This way, the integrity of an arbitrary amount of data can be efficiently assured by arranging this data in a Merkle tree form and securely storing the contents of its root node. 

\begin{figure*}[t]
     \centering
     \subfigure{\label{fig:data_hashing}
     \includegraphics[width=0.45\linewidth]{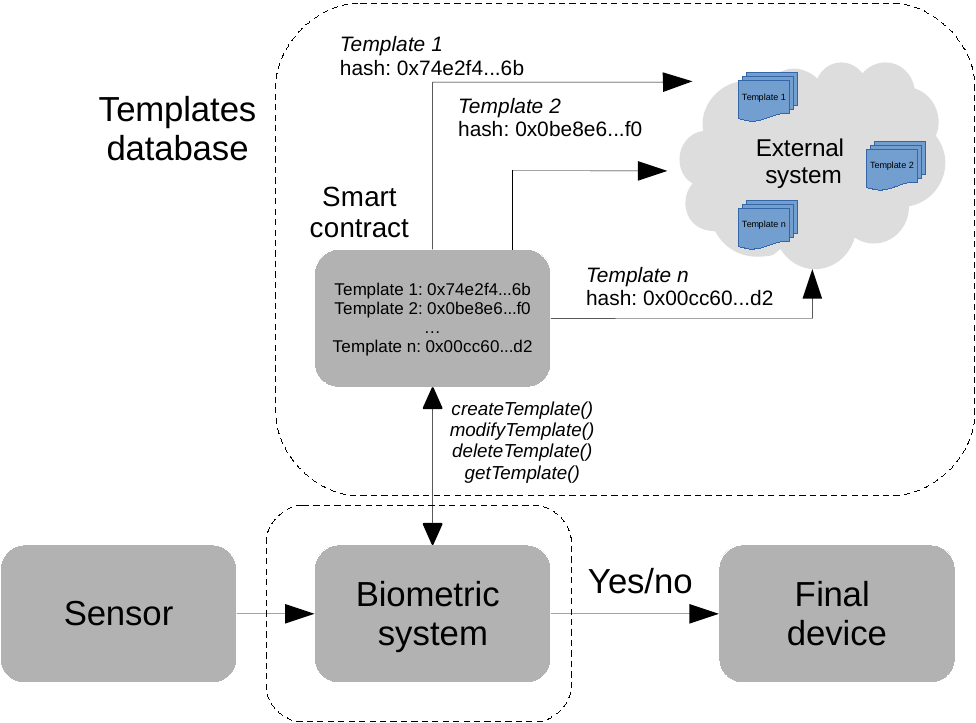}}
     \hspace{0.07\textwidth}
      \subfigure{\label{fig:merkle_tree}
     \includegraphics[width=0.45\linewidth]{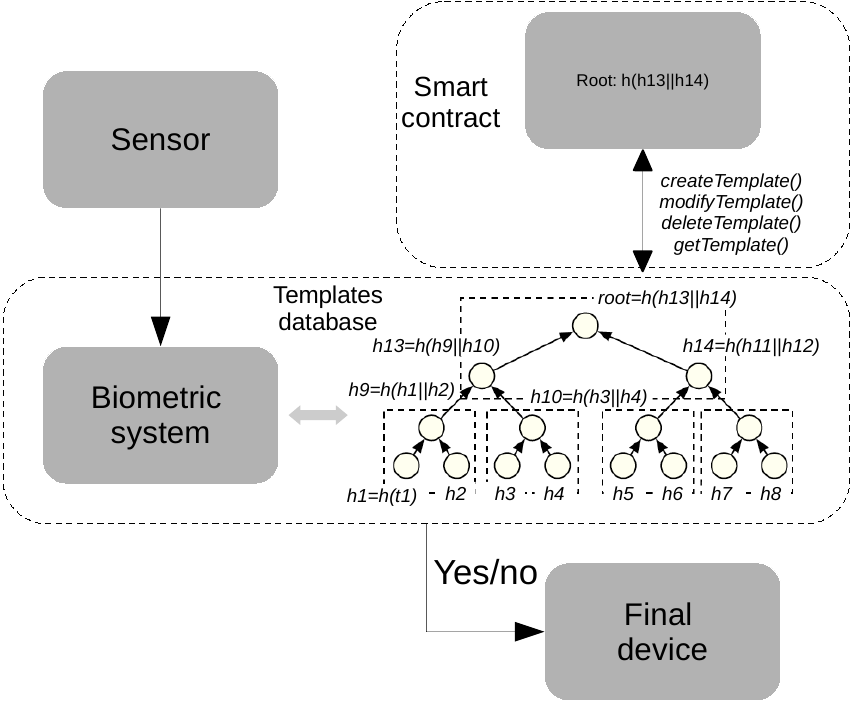}}
     \caption{Biometric systems using data hashing (left) and Merkle trees (right) blockchain storage techniques.}
     \label{fig:hash_merkle_tree}
\end{figure*}

Regarding biometric template protection using blockchains, a biometric system using this technique would maintain a Merkle tree, storing a template at each node and assuring the root node in a smart contract. Therefore, when a new biometric template is created (after the enrollement stage), or an existing one is modified or deleted, the tree is re-calculated and the new root is updated in the blockchain. A simplified scheme of this approach can be found in Fig. \ref{fig:hash_merkle_tree} (right).



\section{Experimental methods}\label{sec:methods}

\subsection{Blockchain technology}\label{sec_approach_blockchain}

The baseline architecture considered for performing the experiments presented in this work was initially introduced in \cite{2019arXiv190305496D}. This architecture substitutes the usual template database of a biometric system by a blockchain, adding basic operations (i.e., creation, modification and deletion of templates) through the use of smart contracts. 

This design provides some advantages:
\begin{itemize}
    \item The modifications to the existing biometric architectures are minimal, so that usual biometric techniques and algorithms (e.g., feature extraction and matching) can be used normally.
    \item Since the biometric process is performed off-chain, this architecture avoids the scalability problems of public blockchains (except in a massive batch of user registration during the system startup, for example). 
    \item No need to use complex smart contracts, which facilitates development and reduces execution costs. Smart contracts do not implement biometric ``logic'', but only the minimum necessary functions to manage the storage of the templates.
\end{itemize}



As stated, we have implemented a basic smart contract, that has been deployed to the Ropsten Ethereum testnet. The contract models a biometric template as a data structure {\small\texttt{BiometricTemplate}} implemented as a raw array of bytes. This structure is stored in a mapping (or hash table), with an identifier number for the user acting as the mapping key {\small\texttt{mapping(uint => BiometricTemplate)}}. The source code of smart contracts can be found in Appendix \ref{sec:smart-contract}. The main operations are described below: 
\begin{itemize}
    \item \textbf{Creation:} Receives the user ID, template data and metadata, and adds a new {\small\texttt{BiometricTemplate}} structure to the blockchain. 
    \item \textbf{Modification:} Modifies the template of an existing user. For a hash table storage scheme, this is equivalent to an addition operation.
    \item \textbf{Deletion:} Removes the link between a specific template and user ID. Due to the public nature of Ethereum, technically the old template data remains forever in the blockchain. 
    \item \textbf{Retrieval:} Retrieves the {\small\texttt{BiometricTemplate}} structure for a user. This function is a \textit{call}, not a transaction as the rest of functions. This operation is usually read-only (and, therefore, free to execute), while the previous three operations were potentially state-changing.
\end{itemize}

\subsection{Biometric systems}\label{sec_approach_biometric_Systems}
Two different biometric traits are considered in the analysis: 1) face, and 2) dynamic signature. In this way we experiment both with image-based physiological biometrics, and with signal-based behavioral biometrics.

\subsubsection{Face biometrics}\label{sec_approach_face_Systems}
One of the most popular face recognition based on deep convolutional neural networks (DCNNs) are evaluated in this study: VGG-Face \cite{vggface}. 



In this system images are propagated through the CNN obtaining the features at the last fully connected layer. The final matching score is computed through the Euclidean distance of the features obtained from each face image. The dimension of the face features are of 4,096.

\subsubsection{Dynamic signature biometrics}\label{sec_approach_signature_Systems}
Two popular approaches are evaluated in this study: \textit{i}) feature-based systems (a.k.a. global systems), and \textit{ii}) time functions-based systems (a.k.a. local systems).

For the global system, we extract for each signature a total of 100 global features from the normalized \textit{X} and \textit{Y} spatial coordinates. These features are described in~\cite{2015_EncyBio_SignFeat}, and are related to time, kinematic, direction, and geometry information. For the similarity computation, the Mahalanobis distance is used to compare the similarity between a signature and a claimed user model. 


 

For the local system, a total of 21 local features are extracted from the normalised signals \textit{X} and \textit{Y} spatial coordinates~\cite{2015_IEEEAccess_InterSign_Tolosana}. For the similarity computation, DTW is used to compare the similarity between genuine and query input samples, finding the optimal elastic match among time sequences that minimises a given distance measure. 




\section{Experimental protocol}\label{sec:protocol_setup}

This section describes the main characteristics of the biometric systems evaluated, and the key tradeoffs involved in the integration of blockchain technology in both \textit{face} and \textit{dynamic signature} biometric traits. 

This integration is evaluated in terms of cost of storage, execution, and performance in the Ethereum blockchain.

Two popular biometric databases are considered for the analysis of face and signature biometrics: Labeled Faces in the Wild (LFW)~\cite{learned2016labeled}, and Biosecure~\cite{ortega2010multiscenario}. 

\subsection{Face}

\subsubsection{Databases}\label{sec:face_databases}

Face verification experiments are conducted on the LFW database (Labeled Faced in the Wild) \cite{lfw}. LFW is one of the most popular datasets used in face recognition with more than 13,000 face images of famous people collected from the web. We have used the aligned dataset where each image was aligned with funneling techniques. 

\subsubsection{Data processing}



The VGG-Face pre-trained model was tested using the \emph{unrestricted} and \emph{outside training data} protocols proposed in \cite{lfw}. The VGG-Face model was trained with VGG-Face database \cite{vggface}, therefore there is not extra training for the pre-trained model used here. The evaluation results are computed for 6,000 one-to-one comparisons composed by 3,000 genuine pairs (pairs of images from the same person) and 3,000 impostor pairs (pairs of images belonging to different persons) following the protocols from LFW database. 



\begin{figure}[t]
     \centering
     \includegraphics[width=1\linewidth]{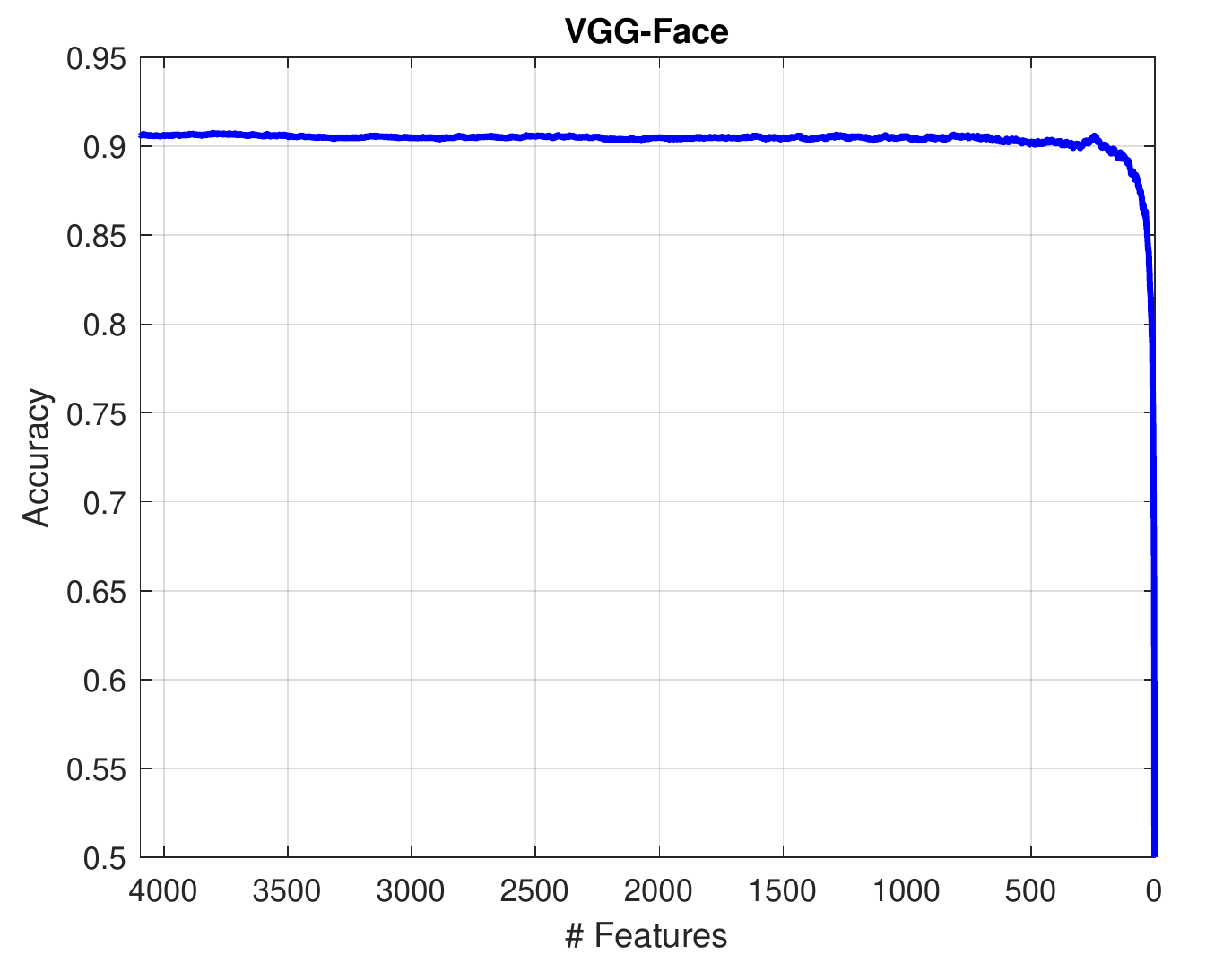}
     \caption{System performance results in terms of the size of the feature embeddings for VGG-Face CNNs model.}
     \label{fig:faceresults}
\end{figure}

\begin{figure*}[t]
     \centering
     \subfigure{\label{fig:global_signature}
     \includegraphics[width=0.43\linewidth]{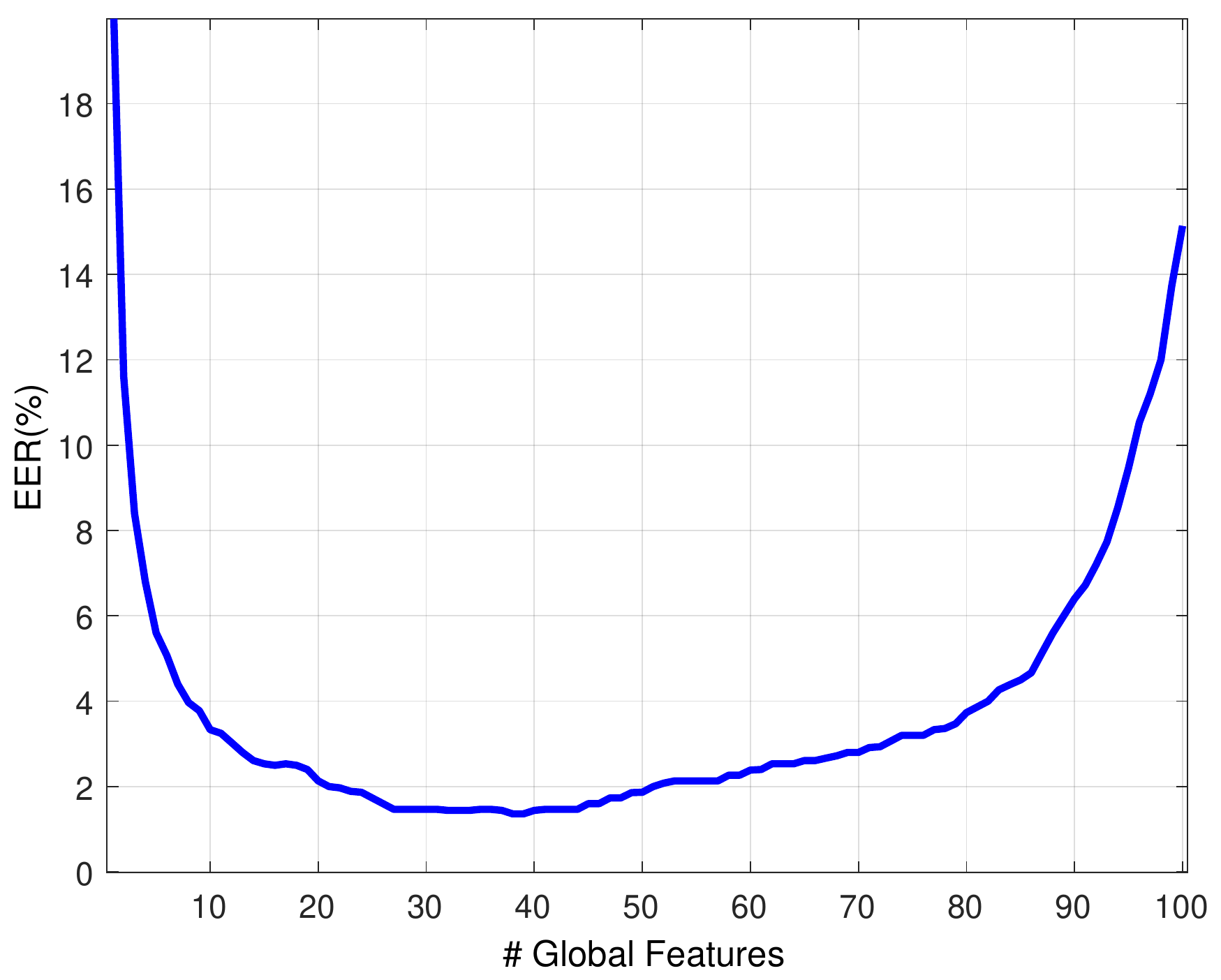}}
      \subfigure{\label{fig:local_signature}
     \includegraphics[width=0.44\linewidth]{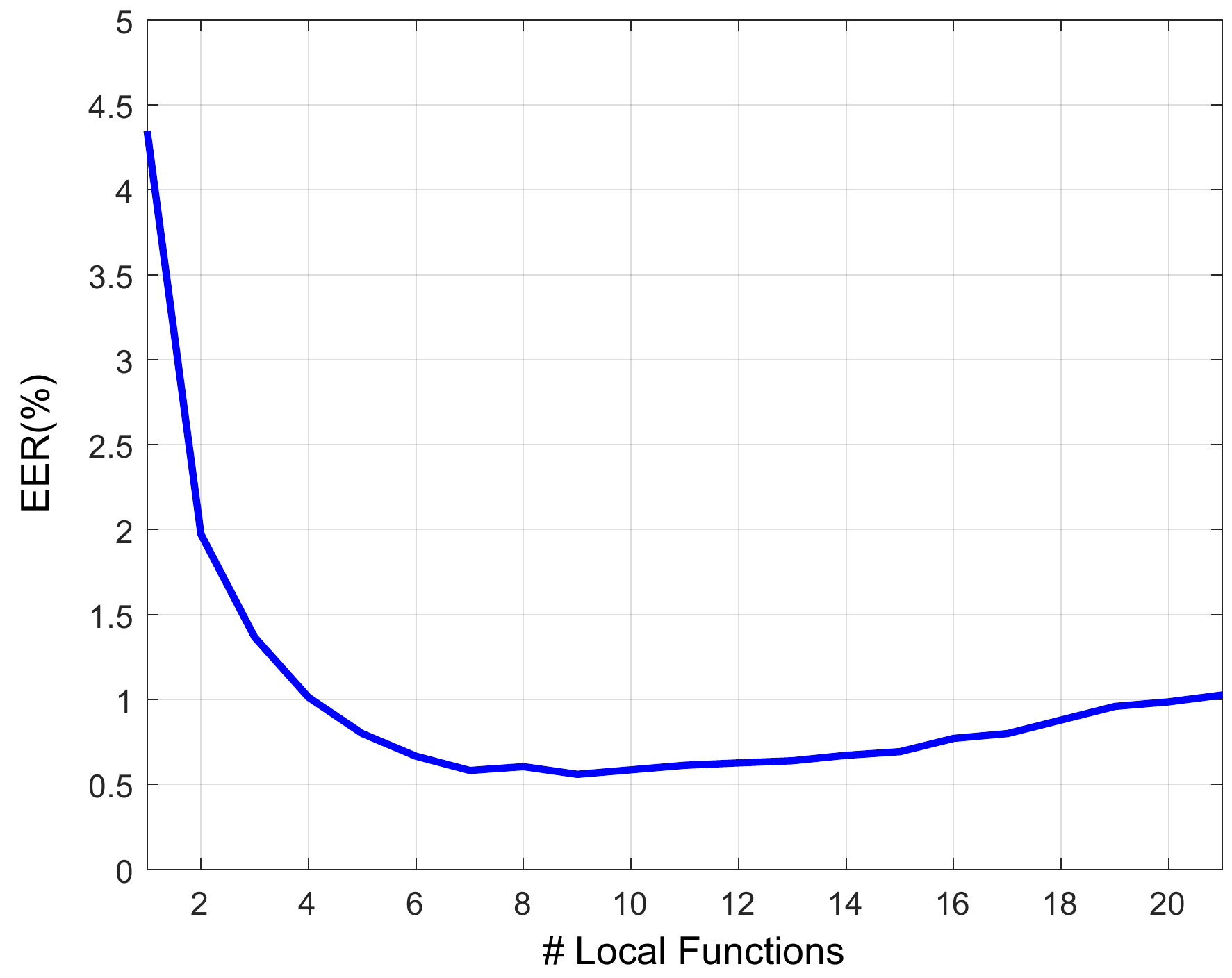}}
     \caption{System performance results in terms of the size of the optimal feature/time function vector selected by the SFFS algorithm. \newline Left: global system. Right: local system. For local system: \#Features = \#Local Functions $\times$ Average time samples per signature (343).}
     \label{fig:signature_size_template}
\end{figure*}

\subsection{Dynamic signature}

\subsubsection{Databases}\label{sec:sig_databases}
 The dynamic signature verification technology is analyzed using the Biosecure DS2 dataset. This dataset was captured using a Wacom Intuous 3 digitizing tablet with an inking pen in an office-like scenario, providing the following information: \textit{X} and \textit{Y} spatial coordinates, pressure, and timestamp (sampling frequency 100 Hz).
 

\subsubsection{Data processing}

In this study, we consider a set of 50 users. For each user, the first 5 genuine signatures of the first session are used for training, whereas the 15 genuine signatures of the second session are left for testing in order to consider the inter-session variability. In this study we analyze the robustness of our proposed system against random (zero-effort) forgeries. Scores are obtained by comparing the training signatures with one genuine signature of the remaining users. For the global system, scores are obtained by comparing signatures against the user model, while for the local system, the average score of the five one-to-one comparisons is used. 

\subsection{Blockchain integration tradeoffs}

\subsubsection{Face}


The system performance results in terms of EER (\%) for VGG-Face CNN model is depicted in Fig.~ \ref{fig:faceresults} for different sizes of the biometric template. This analysis has been carried out by removing features randomly from the original feature embedding.


Analyzing results in Fig. \ref{fig:faceresults}, in general the system performance is very stable while we gradually remove features. VGG-Face is able to obtain a verification rate with an accuracy of 89\% only using 100 features (only 2.5\% from the original 4096 features). This behavior shows that there is a very high redundancy within the feature embedding of CNNs face models, which makes possible to obtain very competitive verification performance while keeping only a small set of features.

\subsubsection{Dynamic signature}

The system performance results in terms of EER (\%) of both global and local systems are depicted in Fig.~\ref{fig:signature_size_template} for different sizes of the biometric template. This analysis has been carried out using  Sequential Forward Floating Search (SFFS) in order to select the best subsets of global and local features that improve the system performance in terms of EER (\%). 

Analyzing in Fig.~\ref{fig:signature_size_template} (left) the global approach, the system performance improves when increasing from 1 to 30-40 global features. After that, a degradation of the system performance is produced when adding more global features to the optimal feature vector. Therefore, in order to reduce the cost of saving the biometric templates in the blockchain platform, and also achieve the best possible system performance, we propose to save the best 30 global features in the biometric template, achieving this way a final 1.5\% EER. 

The same analysis has been carried out for the local approach in Fig.~\ref{fig:signature_size_template} (right). The system performance improves when adding more local features, achieving for the best system performance a final 0.5\% EER using 9 local functions (total number of features = 9 local functions $\times$ average time samples per signature = 3,087 features). 

\begin{table*}[]
\centering
\resizebox{\textwidth}{!}{%
\begin{tabular}{|c|c|c|c|c|c|c|}
\hline
\multicolumn{2}{|c|}{\textbf{Biometric}} & \multirow{2}{*}{\textbf{Operation}} & \multicolumn{3}{c|}{\textbf{Storage scheme}} & \textbf{Performance} \\ \cline{1-2} \cline{4-7} 
\textit{Scheme} & \textit{\begin{tabular}[c]{@{}c@{}}Template \\ size\end{tabular}} &  & \textit{Full on-chain} & \textit{\begin{tabular}[c]{@{}c@{}}Data hashing\\ (cost per template)\end{tabular}} & \textit{\begin{tabular}[c]{@{}c@{}}Merkle trees \\ (cost for any number \\ of templates)\end{tabular}} & \textit{\begin{tabular}[c]{@{}c@{}}Execution time \\ (average)\end{tabular}} \\ \hline
- & - & \begin{tabular}[c]{@{}c@{}}Smart contract \\ deployment\end{tabular} & \multicolumn{3}{c|}{\begin{tabular}[c]{@{}c@{}}498274 gas \\ (\$0.06972)\end{tabular}} & 19.19 secs \\ \hline
\multirow{8}{*}{Signature} & \multirow{4}{*}{\begin{tabular}[c]{@{}c@{}}Global \\ 30 x 16 bits\end{tabular}} & Creation & \multirow{2}{*}{\begin{tabular}[c]{@{}c@{}}108844 gas \\ (\$0.014)\end{tabular}} & \multicolumn{2}{c|}{\multirow{2}{*}{\begin{tabular}[c]{@{}c@{}}86848 gas\\ (\$0.0122)\end{tabular}}} & \multirow{2}{*}{10.66 secs} \\ \cline{3-3}
 &  & Modification &  & \multicolumn{2}{c|}{} &  \\ \cline{3-7} 
 &  & Deletion & \begin{tabular}[c]{@{}c@{}}21378 gas \\ (\$0.003)\end{tabular} & \multicolumn{2}{c|}{\begin{tabular}[c]{@{}c@{}}18850 gas\\ (\$0.0026)\end{tabular}} & 11.55 secs \\ \cline{3-7} 
 &  & Retrieval & - & - & - & - \\ \cline{2-7} 
 & \multirow{4}{*}{\begin{tabular}[c]{@{}c@{}}Local \\ 3087 x 16 bits\end{tabular}} & Creation & \multirow{2}{*}{\begin{tabular}[c]{@{}c@{}}4358990 gas \\ (\$0.610)\end{tabular}} & \multicolumn{2}{c|}{\multirow{2}{*}{\begin{tabular}[c]{@{}c@{}}86848 gas\\ (\$0.0122)\end{tabular}}} & \multirow{2}{*}{12.61 secs} \\ \cline{3-3}
 &  & Modification &  & \multicolumn{2}{c|}{} &  \\ \cline{3-7} 
 &  & Deletion & \begin{tabular}[c]{@{}c@{}}504322 gas \\ (\$0.07)\end{tabular} & \multicolumn{2}{c|}{\begin{tabular}[c]{@{}c@{}}18850 gas\\ (\$0.0026)\end{tabular}} & 12.85 secs \\ \cline{3-7} 
 &  & Retrieval & - & - & - & - \\ \hline
\multirow{4}{*}{Face} & \multirow{4}{*}{\begin{tabular}[c]{@{}c@{}}VGG-Face \\ 100 x 32 bits\end{tabular}} & Creation & \multirow{2}{*}{\begin{tabular}[c]{@{}c@{}}352912 gas \\ (\$0.049)\end{tabular}} & \multicolumn{2}{c|}{\multirow{2}{*}{\begin{tabular}[c]{@{}c@{}}86848 gas\\ (\$0.0122)\end{tabular}}} & \multirow{2}{*}{10.53 secs} \\ \cline{3-3}
 &  & Modification &  & \multicolumn{2}{c|}{} &  \\ \cline{3-7} 
 &  & Deletion & \begin{tabular}[c]{@{}c@{}}49192 gas \\ (\$0.0068)\end{tabular} & \multicolumn{2}{c|}{\begin{tabular}[c]{@{}c@{}}18850 gas\\ (\$0.0026)\end{tabular}} & 16.38 secs \\ \cline{3-7} 
 &  & Retrieval & - & - & - & - \\ \hline
\end{tabular}%
}
\caption{We have considered a gas price of 1 gwei (1 gwei = $10^{-9}$ ETH), and 1 ETH = \$140 (accurate at time of writing, March 2019).}\label{tab:results}
\end{table*}

\section{Experimental results}\label{sec:results}

This section analyzes the results of the evaluation of the integration of the biometric systems previously described in Ethereum, resumed in Table \ref{tab:results}.

The smart contract developed has been written in Solidity language, and deployed to the Ethereum Ropsten testnet at the address {\small\texttt{0x8f737f448de451db9b1c046be7df3b48839673a1}}, where can be verified with any blockchain explorer like Etherscan \cite{smart_contract}. It is a basic contract, which has not been optimised and does not take care of security issues, and should be used only for experimental purposes. 

Table \ref{tab:results} shows the costs of the different operations over the templates (creation, modification,  deletion, and retrieval) in units of gas and US dollars, for the biometric technologies and blockchain storage schemes evaluated. 

The results clearly prove that the most efficient storage scheme is the one based on Merkle trees. In fact, it is the only one capable of storing any amount of data for the same cost. The rest of the schemes would quickly have a prohibitive cost for the number of templates to be stored in a real environment.


For example, protecting a million of templates would cost between \$14,000 and \$610,000 for the signature system, and \$49,000 for VGG Face using the \textit{full on-chain} storage scheme. Clearly this is not a realistic option, discouraged not only in economic terms, but also for security and performance reasons. 

The \textit{data hashing} scheme would improve significantly those figures, because it does not store the data itself, but only a hash that guarantees the integrity. For the same scenario, the cost would be a much more reasonable amount of \$12,200 for all the biometric technologies. 

Finally, the \textit{Merkle trees} scheme would imply a cost of only one cent of dollar (\$0.0122) for the storage of any amount of templates. In addition, also the modification operation of a template would have the same cost. However, even for a biometric system operating in a large corporation or environment, these costs seem reasonable.

Of course, all these prices could vary greatly depending on the price of ether, which, as the rest of cryptocurrencies, usually suffers sharp increases and falls in price. However, because it only needs to store 256 bits regardless of the total volume of data, the Merkle tree scheme would still have a reasonable cost in any case.

In terms of execution time and performance, the experiments also show that this hybrid system is viable. It is important to note that the tests have been carried out in a testnet, where the confirmation times are higher and have greater variability than in the mainnet. Times have been measured performing each operation ten times, discarding the minimum and maximum times, and calculating the average of the rest.

As can be seen, the execution time is slightly higher than 10 seconds for most of the operations and biometric systems, which seems an acceptable time for the usability of the system even during a user enrollment, for example.

Finally, the retrieval operation, necessary for the verification of a template, is a read-only operation and, therefore, free of cost. In addition, it can be also considered immediate in terms of execution time, due to that the request is processed by the local Ethereum node, and it does not reach the network.




 

	
 





\section{Conclusions}\label{sec:conclusions}

In this paper we have explored the viability of biometric systems based on blockchain with focus on storing the biometric templates. This experimental exploration has been around key cost-performance tradeoffs, in particular: time of execution of the transactions, economic cost, and biometric performance.

We have first discussed the main storage schemes for public blockchains (Ethereum), and implemented a smart contract for the estimation of its storage cost. The results obtained prove that straightforward schemes such as the direct storage of the biometric templates on-chain, or direct data hashing, are not appropriate for a real biometric system. However, when Merkle trees are included as an intermediate data structure, the storage costs become fixed regardless the total volume of data to store, and reduced execution times (between 10 - 20 seconds for \textit{write} operations) are obtained. The \textit{read} operations (retrieving) of templates are usually free of cost and very fast to execute, because they are processed locally. 

In brief, in this work we have shown that the integration of biometric and public blockchains is possible both from an economic and performance perspective, including two case studies with state-of-the-art methods and protocols in face and signature biometrics.

\section*{Acknowledgments}
This work has been supported by projects: BIBECA (TEC2018 MINECO), Bio-Guard (Ayudas Fundación BBVA a Equipos de Investigación Científica 2017), UAM Cecabank chair on Biometrics, and UAM-Grant Thornton chair on Blockchain. Ruben Tolosana is supported by a FPU Fellowship from Spanish MECD.

%
{\small
\bibliographystyle{styles/ieee_fullname}
\bibliography{biblio}
}

\onecolumn
\appendix
\newpage
\section{Appendix: Smart contract source code}\label{sec:smart-contract}

{\small
\begin{verbatim}
pragma solidity >=0.4.22 <0.6.0;
contract BioBlockchain {


    /// This struct models a simple biometric template
    struct BiometricTemplate {
        bytes templateMetadata;
        bytes templateData;
    }
    
    /// Each template is indexed by an user ID
    mapping(uint => BiometricTemplate) templates;

    /// Store a new template
    function createNewTemplate(uint _templateID, 
                                bytes memory _templateMetaData, 
                                bytes memory _templateData) public {
        
        /// Add new template to mapping
        templates[_templateID].templateMetadata = _templateMetaData;
        templates[_templateID].templateData = _templateData;
        
    }
    
    /// Return a user template
    function getTemplate(uint _templateID) view public returns (bytes memory) {
        
        return(templates[_templateID].templateData);
        
    }
    
    /// Modify a user template
    function modifyTemplate(uint _userID, 
                            bytes memory _newTemplateMetaData,
                            bytes memory _newTemplateData) public {
        
        // Due to that a mapping is internally implemented using
        // a hash table, the modification operation is equivalent
        // to a insertion
        createNewTemplate(_userID, _newTemplateMetaData, _newTemplateData);
        
    }
    
    /// Return an specific template
    function deleteTemplate(uint _userID) public {
        
        delete templates[_userID];
        
    }

}
\end{verbatim}}

\end{document}